\documentclass[aps,pre,reprint]{revtex4-1}
\usepackage{amsmath}
\usepackage{graphicx}
\usepackage{color}
\usepackage{amssymb}
\usepackage{wasysym}
\usepackage{soul}
\input{epsf}

\begin{document}

\bibliographystyle{prsty}

\title{Depletion-induced biaxial nematic states of boardlike particles}

\author{S. Belli$^1$, M. Dijkstra$^2$ and R. van Roij$^1$}

\address{$^1$Institute for Theoretical Physics, Utrecht University, Leuvenlaan 4, 3584 CE Utrecht, The Netherlands}
\address{$^2$Soft Condensed Matter Group, Debye Institute for NanoMaterials Science, Utrecht University, Princetonplein 5, 3584 CC Utrecht, The Netherlands}

\begin{abstract}

With the aim of investigating the stability conditions of biaxial nematic liquid crystals, we study the effect of adding a non-adsorbing ideal depletant on the phase behavior of colloidal hard boardlike particles. We take into account the presence of the depletant by introducing an effective depletion attraction between a pair of boardlike particles. At fixed depletant fugacity, the stable liquid crystal phase is determined through a mean-field theory with restricted orientations. Interestingly, we predict that for slightly elongated boardlike particles a critical depletant density exists, where the system undergoes a direct transition from an isotropic liquid to a biaxial nematic phase. As a consequence, by tuning the depletant density, an easy experimental control parameter, one can stabilize states of high biaxial nematic order even when these states are unstable for pure systems of boardlike particles.     

\medskip


\medskip

\noindent e-mail: s.belli@uu.nl

\end{abstract}

\maketitle

\section{Introduction}
\label{sec:intro}

Onsager's intuition that purely repulsive rods undergo an entropy-driven transition from an isotropic ($I$) to an orientationally ordered nematic ($N$) phase constitutes one of the major milestones in our understanding of liquid crystals \cite{onsager}. The key ingredient of this phenomenon relies on considering markedly non-spherical particles, which can be modeled as cylindrically symmetric ``rods'' and ``plates''.  In the early 1970s Freiser pointed out that a richer phase behavior is expected, if the assumption of cylindrical symmetry is released  \cite{freiser}. Besides the usual prolate ($N_+$) and oblate ($N_-$) uniaxial nematic phases, normally developed by uniaxial rods and plates, respectively, a novel nematic phase with an increased orientational order can appear in the phase diagram. Such a liquid crystal phase is characterized by alignment along three directors and, consequently, by the presence of two distinct optical axes, hence the name {\em biaxial nematic} ($N_B$) \cite{tschierske}. Further studies suggested that $N_B$ stability could be interpreted as a balanced competition between rodlike (favoring $N_+$) and platelike (favoring $N_-$) behavior \cite{alben, straley, mulder}. 

In more than 40 years since its first theoretical prediction, extensive theoretical \cite{boccara,mulder,taylor,vanroij,martinez2,varga,vanakaras,dematteis,longa,allender,martinez,belli} and simulation \cite{allen,camp,berardi2,bates3,bates2,berardi,cuetos} work has been devoted to identify the conditions under which a stable $N_B$ phase could be observed. The practical limitations in this sense are testified by the fact that, apart from the micellar system studied by Yu and Saupe \cite{yu}, no such state has been observed for more than 30 years. A renewed interest towards the topic has grown due to the first experimental realization of thermotropic $N_B$ liquid crystals in systems of bent-core molecules a few years ago  \cite{madsen,acharya}. In lyotropics, a remarkably stable $N_B$ phase was recently discovered in a colloidal suspension of mineral boardlike particles \cite{vandenpol}.

Boardlike particles, that is, particles with the symmetry of a brick, represent the simplest model in which an $N_B$ phase has been predicted \cite{tschierske}. However, the emergence of smectic layering is expected to prevent the realization of this phase, unless the constituent particles are designed with a precision far beyond present-day ability \cite{taylor, vandenpol}. A higher $N_B$ stability can be achieved by considering size-polydisperse systems of boardlike particles, as demonstrated by a recent experiment \cite{vandenpol}. In fact, polydispersity seems to enhance $N_B$ stability through two distinct phenomena: (i) a reduced smectic stability \cite{vanakaras} and (ii) an $N_+$-$N_-$ competition, which manifests itself exclusively in systems of slightly elongated (rodlike) boards \cite{belli}. The first phenomenon does not come as a surprise \cite{vanakaras}, since it is well known that polydispersity renders the establishment of long-distance positional ordering unfavorable \cite{barrat, mcrae, bates}. On the contrary, the reason behind the second phenomenon appears to be more obscure.

In this paper we investigate the effect of a non-adsorbing depletant on the biaxial nematic stability of (monodisperse) boardlike particles. Our understanding of depletion dates back to the pioneering work by Asakura and Osawa \cite{asakura} and Vrij \cite{vrij}, who showed that the addition of small co-solutes (e.g. polymers, surfactants, micelles) to a colloidal suspension gives rise to an effective attraction between colloidal particles. Since then, the concepts related to depletion have been widely applied to various scientific fields \cite{tuinier}: in biology by interpreting phenomena like macromolecular crowding \cite{odijk} and protein crystallization \cite{piazza}; in nanotechnology through e.g. the development of self-assembly processes as key-lock structures \cite{odriozola, sacanna}; in condensed matter physics, furnishing answers to fundamental problems like the condition for gas-liquid phase separation \cite{gast}, the kinetics of crystallization \cite{auer, anderson} and the nature of glassy states \cite{weeks}. More recently, the liquid crystal phase behavior of non-spherical colloids, typically rods \cite{lekkerkerker, buitenhuis, bolhuis, dogic, savenko, jungblut} and plates \cite{bates4, vanderkooij, zhang, kleshchanok}, in presence of a depletant has also been addressed. As a general feature, the addition of a depletant reduces the stability of liquid-crystal phases, leading to a direct isotropic-crystal transition at high enough depletant mole fraction. Moreover, when the size of the depletant particles is big enough, one or more critical points appear in the phase diagram, indicating a liquid-gas separation between phases with same spatial symmetries.

In contrast to the aforementioned work on rods and plates, we focus here on the low depletant density limit, where the stability of the nematic liquid crystal phases developed by the pure system of boardlike particles is preserved.
In the same spirit as the Asakura-Oosawa-Vrij model for spheres \cite{asakura,vrij}, we consider the limit of low depletant density and neglect depletant-depletant interactions. For the sake of convenience, we model the depletants as cubic particles excluded from the surface of the cuboids via a hard-core interaction. A mean-field theory at second virial order \cite{onsager, evans} with restricted orientations (Zwanzig model) \cite{zwanzig} constitutes our theoretical framework. The degrees of freedom of the depletant in the partition function can be systematically integrated out, giving rise to an effective potential between boardlike particles \cite{dijkstra, dijkstra2}, where only two-body interactions are considered. The assumption of ideal depletant allows to determine an explicit expression for such a pairwise depletion potential. We show that, by varying the depletant density, the system develops an $N_+$-$N_-$ competition remarkably similar to that predicted for a polydisperse system of boardlike particles in absence of depletant \cite{belli}. If in Ref. \cite{belli} the origin of this competition is not evident, here it appears to be due to a balance between the hard-core repulsion between boardlike particles, favoring $N_+$ ordering, and the depletion attraction, favoring $N_-$ ordering. As a consequence of this effect, the biaxial nematic phase appears to be stable over a wide range of depletant density. We therefore suggest that the concentration of a non-adsorbing depletant furnishes in practical situations the simplest, though effective, way to control the liquid-crystal phase behavior of boardlike particles and to select states of high biaxial-nematic stability.

The paper is organized as follows. We illustrate in Sec. \ref{sec:theory} our theoretical framework and in Sec. \ref{sec:model} the model describing the boards-depletant mixture. Sec. \ref{sec:results} is devoted to the results, whereas in Sec. \ref{sec:conclusions} we draw our conclusions.  

\section{Second-virial density functional theory with restricted orientations}
\label{sec:theory}

\begin{figure}
\includegraphics[scale=0.14]{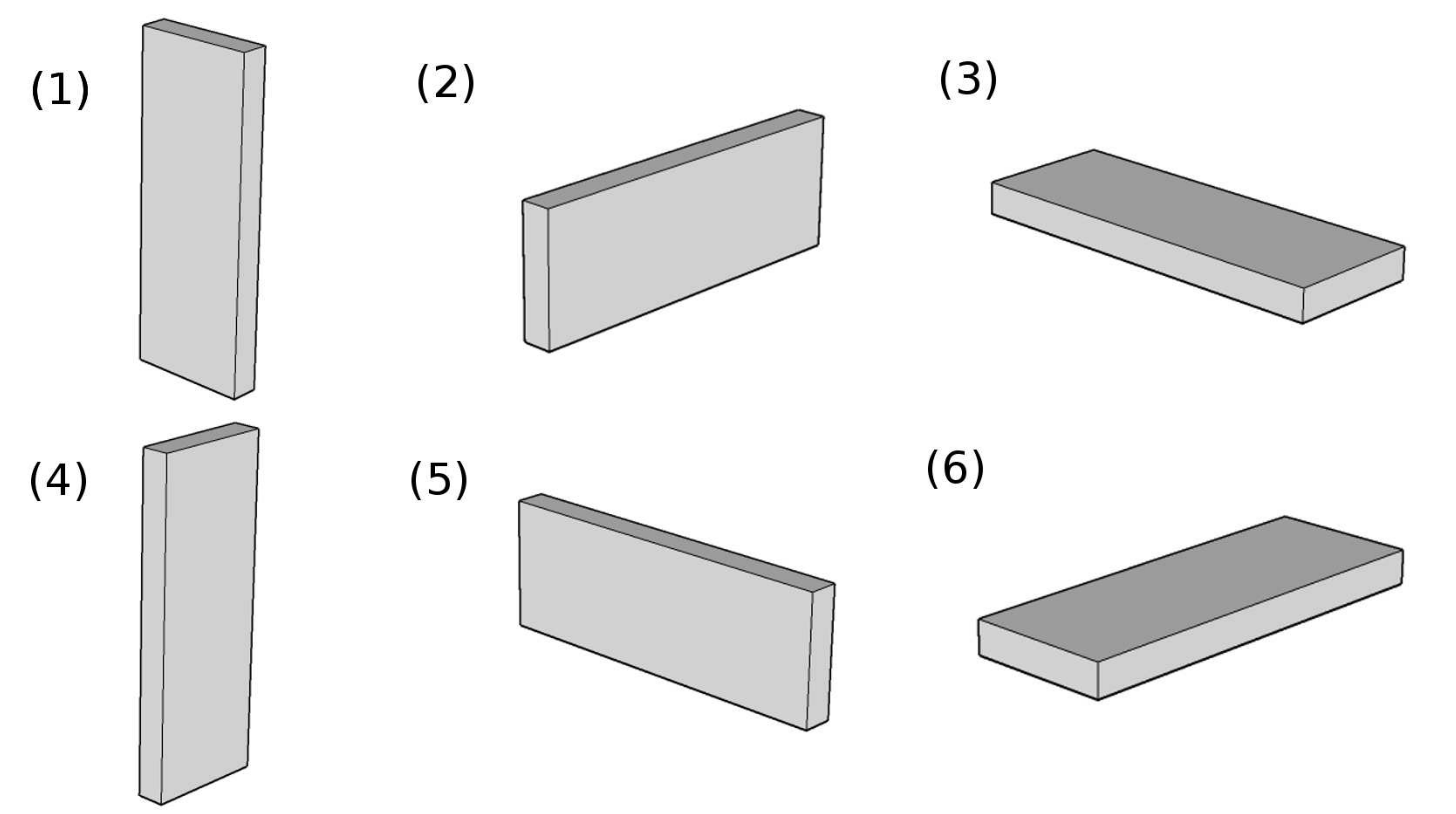}
\caption{\label{fig1} The 6 independent orientations of a boardlike particle within the restricted orientations (Zwanzig) model.} 
\end{figure}

We consider a system of $N$ boardlike particles with dimensions $l \times w \times t$ ($l > w > t$ and particle volume $v=l w t$) in a box of volume $V$ at temperature $T$. Accounting for the orientational degrees of freedom at the single-particle level requires a numerically demanding description based on $3$ Euler angles. In order to circumvent this problem while keeping the essential physics of the system, we turn to the so-called Zwanzig model: the only allowed orientations are those with the main particle axes aligned along the axes of a fixed reference frame \cite{zwanzig}. Within this model a boardlike particle can take the $6$ orientations depicted in Fig. \ref{fig1}, and the orientation distribution function (ODF) is a $6$-dimensional vector $\boldsymbol{\psi}$ with components $\psi_i$ ($i=1,...,6$), subject to the normalization condition

\begin{equation}
\sum_{i=1}^6 \psi_i = 1.
\label{eq1} 
\end{equation}

Being interested in the low-density phase behavior of the system, where the stable phases are expected to be homogeneous in space, we can neglect spatial modulations in the single-particle density. Under these conditions, the intrinsic free energy $\mathcal{F}$ reads \cite{belli}

\begin{equation}
\frac{\beta \mathcal{F}[\boldsymbol{\psi}]}{N} = \ln(\eta) + \sum_{i=1}^6 \psi_i \ln(\psi_i ) + \frac{\beta \mathcal{F}_{\mathrm{exc}}[\boldsymbol{\psi}]}{N},
\label{eq2} 
\end{equation}
where $\beta=(k_B T)^{-1}$, $k_B$ is the Boltzmann constant and $\eta=N v/V$ is the packing fraction. A closed expression for the excess free energy $\mathcal{F}_{\mathrm{exc}}$ in terms of the ODF is not known in general, but for short-range pairwise additive potentials it is possible to write it as a virial series in $\eta$. Let $u_{i i'}(\mathbf{r})$ be the interaction potential between a pair of particles with orientations $i$ and $i'$, respectively, and a separation $\mathbf{r}$ between their centers of mass. At second virial order the excess free energy reads

\begin{equation}
\frac{\beta \mathcal{F}_{\mathrm{exc}}[\boldsymbol{\psi}]}{N} = \frac{\eta}{2 v} \sum_{i,i'=1}^{6} E_{i i'} \psi_{i} \psi_{i'},
\label{eq3} 
\end{equation} 
where the second-virial coefficients $E_{i i'}/2$ are given by

\begin{equation}
E_{i i'} = -\int_V d \mathbf{r} f_{i i'}(\mathbf{r}),
\label{eq4} 
\end{equation}
with the Mayer function

\begin{equation}
f_{i i'}(\mathbf{r}) = \exp \bigl[-\beta u_{ii'}(\mathbf{r}) \bigr] - 1. 
\label{eq5}
\end{equation}
More refined approximations than Eq. (\ref{eq3}) have been recently developed in order to take into account higher-order virial terms, which would give a quantitatively more reliable description \cite{martinez}. On the other hand, truncating the virial series at second order allows for an appreciable simplification of the mathematics and the numerics involved, while retaining the essential physics.

Let us indicate with $X_i$ the main axis ($l$, $w$ or $t$) of a particle with orientation $i$ along the $x$ axis of a fixed reference frame, and similarly with $Y_i$ and $Z_i$. Within the Zwanzig model each of the $6$ independent orientations of a particle can be identified by $(X_i,Y_i,Z_i)$, which is one of the $6$ permutations of the three elements $l$, $w$ and $t$. With these definitions one can write the interaction potential between two identical boardlike particles, modeled as hard cuboids, as

\begin{equation}
\beta u_{i i'}(\mathbf{r}) =
\begin{cases}
\infty \hspace{0.8cm} \text{if $|x|< (X_{i}+X_{i'})/2$ } \\
\hspace{1.15 cm} \text {and $|y|<(Y_{i}+Y_{i'})/2$} \\
\vspace{0.15 cm}
\hspace{1.15 cm} \text{and $|z|<(Z_{i}+Z_{i'})/2$;}\\
0 \hspace{1cm} \text{otherwise,}
\end{cases} 
\label{eq6}
\end{equation}  
from which explicit expressions for $E_{ii'}$ in terms of $l$, $w$ and $t$ directly follow through Eqs. (\ref{eq4})-(\ref{eq5}).

At second-virial order the equilibrium ODF at fixed packing fraction $\eta$ is obtained by minimizing the free energy of Eqs. (\ref{eq2})-(\ref{eq3}) with respect to $\boldsymbol{\psi}$, subject to the normalization condition Eq. (\ref{eq1}). In practice, the minimization problem is performed by numerically solving the system of $6$ non-linear Euler-Lagrange equations

\begin{equation}
\psi_i = C \exp \Biggl[-\frac{\eta}{v} \sum_{i'=1}^{6} E_{i i'} \psi_{i'} \Biggr], 
\label{eq7}
\end{equation}
with the proportionality constant $C$ determined by Eq. (\ref{eq1}). The symmetry of the solution of Eq. (\ref{eq7}) allows to identify the stable homogeneous phase. When $\psi_i=1/6$ for every $i=1,...,6$, the phase is isotropic ($I$), whereas in the opposite case, when all the $\psi_i$ assume different values, the ODF describes a biaxial nematic ($N_B$) phase. When the system is characterized by the presence of a single axis of symmetry (uniaxial nematic phase), the coefficients $\psi_i$ are coupled two-by-two. Let us suppose this axis of symmetry to be the vertical axis of Fig. \ref{fig1}. In this case, we distinguish between prolate uniaxial nematic phase ($N_+$), when the most likely configurations of Fig. \ref{fig1} are (1) and (4), and oblate uniaxial nematic phase ($N_-$), when the most likely configurations are (3) and (6).

In the treatment described so far, we assume the system to be homogeneous in space. In order to estimate the limit of validity of this assumption, we adopt bifurcation theory to calculate the minimum packing fraction $\bar{\eta}$, beyond which homogeneous phases are unstable with respect to smectic states \cite{mulder2}. The mathematical details regarding the application of bifurcation theory to the free energy of Eqs. (\ref{eq2})-(\ref{eq3}) can be found in the Supplemental Material of Ref. \cite{belli}, and we report here only the final result. Let us indicate with $Q_{ii'}^{(x)}(q_x)$ the function

\begin{equation}
Q_{ii'}^{(x)}(q_x) = \frac{\eta}{v} \sqrt{\psi_i \psi_{i'}} \int_V d \mathbf{r} \, f_{ii'}(\mathbf{r}) \exp(-i q_x x), 
\label{eq8}
\end{equation}  
where $\boldsymbol{\psi}$ is the ODF of the equilibrium homogeneous phase at packing fraction $\eta$, and analogously for $Q_{ii'}^{(y)}(q_y)$ and $Q_{ii'}^{(z)}(q_z)$. The bifurcation packing fraction $\bar{\eta}_x$ for smectic fluctuations along the $x$ axis is found as the minimum packing fraction at which the $6 \times 6$ matrix with entries $Q_{ii'}^{(x)}(q_x)$ has an eigenvalue $1$ for some $\bar{q}_x$. Therefore, the smectic bifurcation packing fraction is $\bar{\eta}=\min(\bar{\eta}_x, \bar{\eta}_y, \bar{\eta}_z )$. As a final remark, it is important to notice that the present bifurcation analysis allows only to predict when homogeneous phases are unstable with respect to one-dimensional modulations in the single-particle density. Therefore, nothing ensures the corresponding stable inhomogeneous phase to be characterized by one-dimensional (smectic), rather than two- (columnar) or three-dimensional (crystal) positional ordering.

\section{Effective depletion interaction}
\label{sec:model}

Our aim is to study the influence of a depletant on the phase behavior of a system of boardlike particles. Hence, the system described in Sec. \ref{sec:theory} is modified by the addition of a second species of particles (the depletant), modeled as cubes with dimensions $d \times d \times d$. The binary mixture of boardlike particles and depletant is assumed to be in equilibrium with a reservoir of depletant particles at fixed fugacity $z_D = \exp(\beta \mu_D)/\Lambda_D^3$, where $\mu_D$ is the chemical potential of the depletant and $\Lambda_D$ its thermal wavelength. Following the pioneering approaches to the topic \cite{asakura,vrij}, we neglect interactions between depletants, in which case the fugacity $z_D$ coincides with the density $n_D$ in the reservoir. The ideal-depletant assumption is justified {\em a posteriori} by the low packing fractions $n_D d^3$ considered. Modeling the depletant with cubic particles appears to be rather unrealistic, especially if compared to typical polymeric depletants, usually treated as spheres. However, we claim that our choice contains the essential features of the physical phenomenon, while considerably simplifying the mathematics that follows. In the next section we show that the peculiar phase behavior of our system is due to the asphericity of the depletion volume, which, in turns, is a consequence of the asphericity of boardlike particles. Therefore, we do not expect the specific shape of the depletion region (cuboidal for cubic depletant, spherocuboidal for spherical depletant) to play a major role in our results. Moreover, the relative difference between cuboidal and spherocuboidal depletion volume for the values of the particles dimensions considered here amounts to few percentage points. The interactions in the mixture are given by the cuboid-cuboid potential Eq. (\ref{eq6}) between boardlike particles, and by the cuboid-cube potential between boardlike particles and depletant, given by

\begin{equation}
\beta v_{i}(\mathbf{r}) =
\begin{cases}
\infty \hspace{0.8cm} \text{if $|x|< (X_{i}+d)/2$ } \\
\hspace{1.15 cm} \text {and $|y|<(Y_{i}+d)/2$} \\
\vspace{0.15 cm}
\hspace{1.15 cm} \text{and $|z|<(Z_{i}+d)/2$;}\\
0 \hspace{1cm} \text{otherwise,}
\end{cases} 
\label{eq9}
\end{equation}
which explicitly depends on the orientation $i$ of the boardlike particle.


At fixed fugacity $z_D$ the configurational entropy of the depletant is maximized when the total depletion volume, i.e. the region of space forbidden to the depletant due to the presence of boardlike particles, is minimized. As a consequence, an effective attraction between boardlike particles appears. Such a depletion interaction can be explicitly calculated by integrating out the depletant degrees of freedom and must be expressed in general as a sum of many-body interaction terms \cite{dijkstra, dijkstra2}. For the sake of simplicity, we include the effect of the depletant by considering only the effective two-body interaction potential, while neglecting higher order terms. The effective pairwise depletion potential $w_{i i'}(\mathbf{r})$ between cuboids with orientations $i$ and $i'$, respectively, and center-to-center separation $\mathbf{r}$ is given by \cite{gotzelmann}

\begin{equation}
\beta w_{i i'}(\mathbf{r}) = -n_D \mathcal{V}_{i i'}(\mathbf{r}),
\label{eq10}
\end{equation}
with $\mathcal{V}_{i i'}(\mathbf{r})$ the overlap volume of the depletion regions,

\begin{equation}
\mathcal{V}_{i i'}(\mathbf{r})=
\begin{cases}
0 \hspace{1.1cm} \text{if $|x|>(2d+X_{i}+X_{i'})/2$} \\
\hspace{1.3 cm} \text {and $|y|>(2d+Y_{i}+Y_{i'})/2$} \\
\vspace{0.15 cm}
\hspace{1.3 cm} \text{and $|z|>(2d+Z_{i}+Z_{i'})/2$;}\\
\lambda^{(x)}_{i i'} \lambda^{(y)}_{i i'} \lambda^{(z)}_{i i'} \hspace{1cm} \text{otherwise.}
\end{cases} 
\label{eq11}
\end{equation}
Here $\lambda^{(x)}_{i i'}$ is defined as

\begin{equation}
\lambda^{(x)}_{i i'}(x)=
\begin{cases}
d + \frac{X_{i}+X_{i'}}{2} - |x| \hspace{0.5cm} \text{if $|x|>\frac{|X_{i}-X_{i'}|}{2}$} \\
\vspace{0.15 cm} 
\hspace{3 cm} \text{and $|x|<(d+\frac{X_{i}+X_{i'}}{2})$;} \\
d + \min(X_{i},X_{i'}) \hspace{0.5cm} \text{if $|x|<\frac{|X_{i}-X_{i'}|}{2}$,}
\end{cases} 
\label{eq12}
\end{equation}
and analogous definitions hold for $\lambda^{(y)}_{i i'}(y)$ and $\lambda^{(z)}_{i i'}(z)$.

Let us indicate with a tilde the properties obtained by adding the effective two-body depletion potential $w_{i i'}(\mathbf{r})$ to the cuboid-cuboid potential $u_{i i'}(\mathbf{r})$. The Mayer function Eq. (\ref{eq5}) becomes

\begin{equation}
\widetilde{f}_{i i'}(\mathbf{r}) = \exp \bigl[-\beta u_{ii'}(\mathbf{r}) + n_D \mathcal{V}_{i i'}(\mathbf{r}) \bigr] - 1.  
\label{eq13}
\end{equation}
The phase behavior of this effective one-component system can then be calculated by following the prescriptions of Sec. \ref{sec:theory}, with the function $f_{i i'}(\mathbf{r})$ substituted by $\widetilde{f}_{i i'}(\mathbf{r})$. Unfortunately, the expression of $\mathcal{V}_{i i'}(\mathbf{r})$ given in Eq. (\ref{eq11}) does not allow for an analytical calculation of the integrals $\widetilde{E}_{ii'}$ and $\widetilde{Q}_{ii'}^{(x)}(q_x)$ in Eqs. (\ref{eq4}) and (\ref{eq8}). However, an  analytical expression can be obtained by truncating the Taylor series of the Mayer function Eq. (\ref{eq13}) in $n_D \mathcal{V}_{i i'}(\mathbf{r})$,

\begin{equation}
\widetilde{f}_{ii'}(\mathbf{r}) = f_{ii'}(\mathbf{r}) + \sum_{m=1}^{\infty} \frac{n_D^m}{m!} \bigl (\mathcal{V}_{i i'}(\mathbf{r}) \bigr)^m \exp \bigl [-\beta u_{ii'}(\mathbf{r})\bigr]. 
\label{eq14}
\end{equation}
By inserting Eq. (\ref{eq14}) into Eq. (\ref{eq4}), one obtains for the effective excluded-volume coefficients

\begin{equation}
\widetilde{E}_{i i'} = E_{i i'} - \sum_{m=1}^{\infty} \frac{n_D^m}{m!} \int_V d \mathbf{r} \bigl (\mathcal{V}_{i i'}(\mathbf{r}) \bigr)^m \exp \bigl [-\beta u_{ii'}(\mathbf{r})\bigr],
\label{eq15} 
\end{equation}
where the integrals of the r.h.s. can now be solved analytically for every $m$. Similar considerations hold for the functions $\widetilde{Q}_{ii'}^{(x)}(q_x)$ of Eq. (\ref{eq8}). We verified by comparison with exact numerical calculations of the effective excluded-volume coefficients that quantitative agreement can be obtained by truncating the series of Eq. (\ref{eq15}) at fifth order in $n_D$ for all $n_D$ considered in this paper. For consistency, the Taylor expansion in $n_D$ of the functions $\widetilde{Q}_{ii'}^{(x)}(q_x)$ is truncated at the same order. 

\section{Results}
\label{sec:results}

The framework developed in Sec. \ref{sec:model} allows to determine the effective excluded-volume coefficients $\widetilde{E}_{i i'}$ of a system of cuboidal $l \times w \times t$ particles due to the presence of a cubic $d \times d \times d$ depletant at fugacity $z_D$ (and density $n_D=z_D$). The phase behavior of this effective one-component system of boardlike particles is then analyzed by applying the theory described in Sec. \ref{sec:theory}.

\begin{figure}
\includegraphics[scale=0.65]{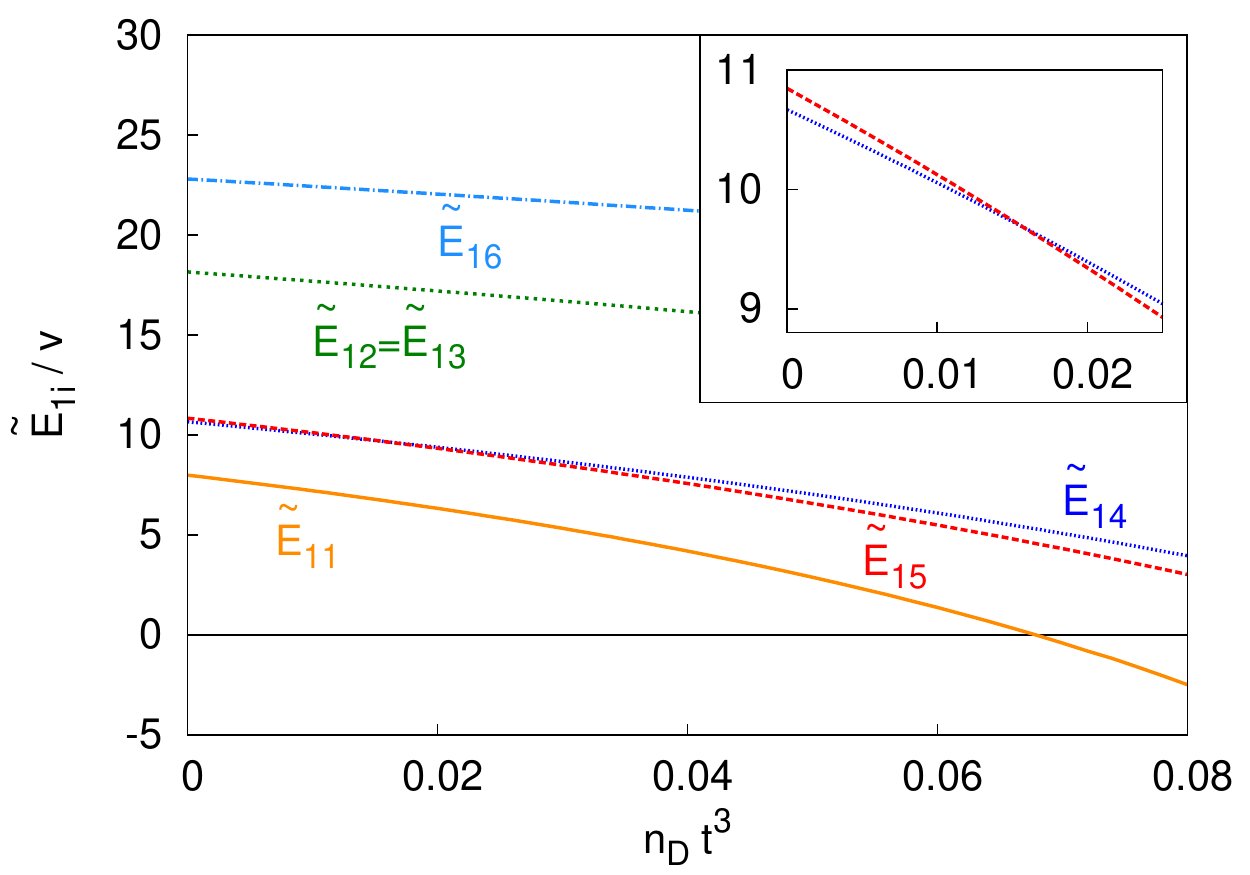}
\caption{\label{fig2} Effective excluded-volume coefficients $\widetilde{E}_{1 i}$ (in units of boardlike particle volume $v=l w t$) for the 6 independent orientational configurations of a pair of boardlike particles in the Zwanzig model as a function of the depletant number density $n_D$. Here the boardlike particles have dimensions $l/t=9.3$, $w/t=3.0$ and are in contact with a reservoir of ideal cubic depletants with side $d/t=1.0$ and number density $n_D$.} 
\end{figure}

It is readily understood from Eq. (\ref{eq15}) that adding the depletion attraction Eq. (\ref{eq10}) to the cuboid-cuboid pairwise potential $u_{ii'}({\mathbf r})$ gives rise to a monotonic decrease of the coefficients $\widetilde{E}_{i i'}$ with $n_D$. This effect is depicted in Fig. \ref{fig2}, where we report the 6 independent values of the matrix elements $\widetilde{E}_{1 i}$, corresponding to the 6 two-particle configurations (1,1), (1,2), (1,3), (1,4), (1,5) and (1,6) (cf. Fig. \ref{fig1}), as a function of the reservoir depletant concentration $n_D$. In order to allow for a comparison with previous experimental \cite{vandenpol} and theoretical \cite{belli} work on the subject, the aspect ratios are chosen as $l/t=9.3$ and $w/t=3.0$, while for the cubic depletant we set $d/t=1.0$. At $n_D=0$ the 6 excluded-volume matrix elements are positive definite, but with increasing $n_D$ their value decreases until becoming negative (see $\widetilde{E}_{11}$ in Fig. \ref{fig1}). Such a behavior is well known from the study of systems of spherically symmetric particles with short-range attractive potentials, where one can define a temperature at which the second virial coefficient changes its sign (``Boyle temperature''). The change in sign of the second-virial coefficient is related to a tendency of the system to develop a gas-liquid phase separation. Also in the present case, where the role of the (inverse) temperature is played by the depletant density $n_D$, this change in sign can indicate a tendency towards a phase separation between two homogeneous phases. On the other hand, when the dimension of the depletant is sufficiently small, one expects the gas-liquid phase separation to be metastable with respect to a broad gas-solid coexistence \cite{dijkstra,dijkstra2,bolhuis,savenko}. As we ignore the stability of inhomogeneous phases like smectic, columnar or crystal states, in the present work we limit our investigations to values of $n_D$ small enough as to guarantee a positive value of all the effective excluded-volume matrix elements, and to avoid strong tendency towards a broad phase separation.

\begin{figure}
\includegraphics[scale=0.65]{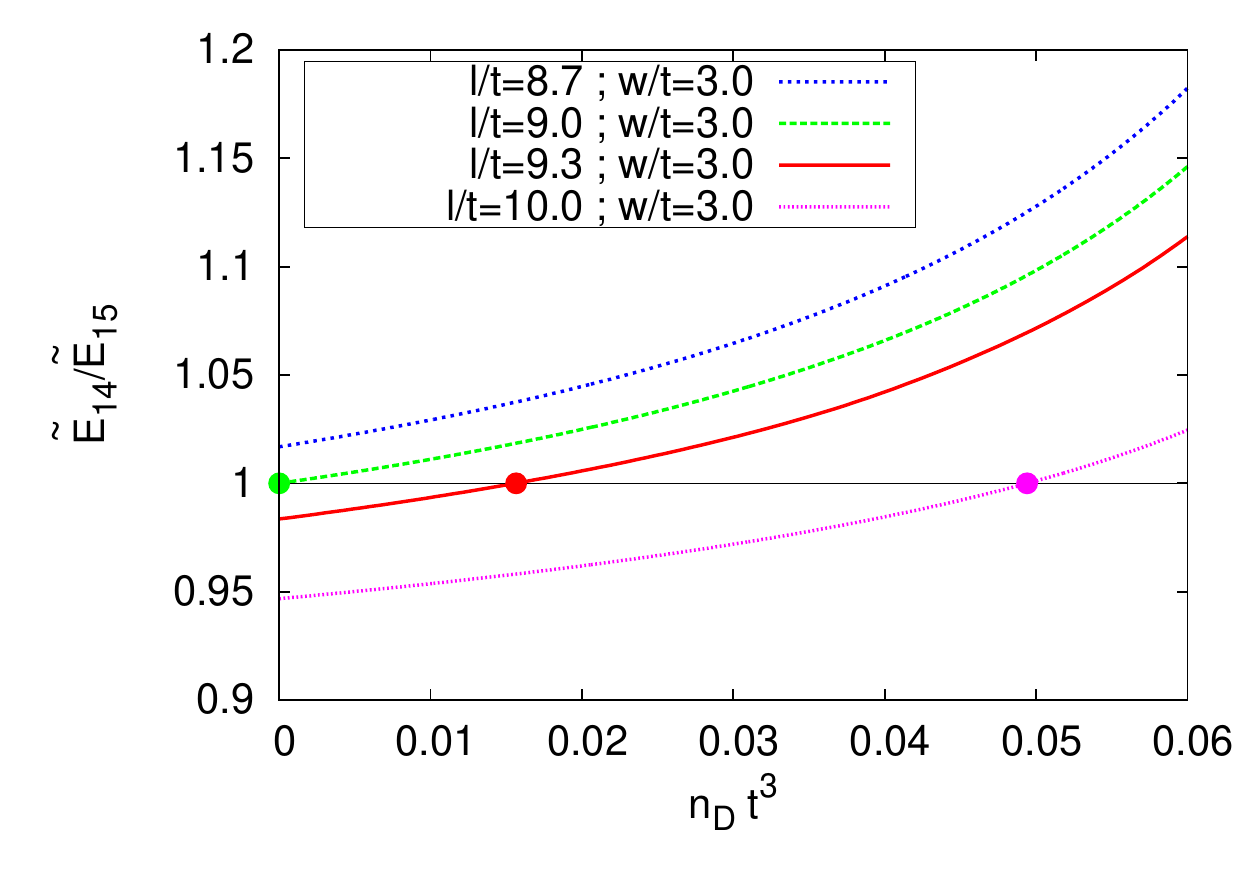}
\caption{\label{fig3} Ratio between the second-virial coefficients corresponding to the two-particle configurations (1,4) and (1,5) of Fig. \ref{fig1} as a function of the depletant density $n_D$ for boardlike particles with $w/t=3.0$ and $l/t=8.7$ ($\nu=l/w-w/t=-0.1$), $l/t=9.0$ ($\nu=0.0$), $l/t=9.3$ ($\nu=0.1$), and $l/t=10.0$ ($\nu=0.33$). The solid circles highlight the value of the depletant density $n_D^*$, defined by the condition $\widetilde{E}_{14}=\widetilde{E}_{15}$.} 
\end{figure}

Although the monotonic decrease with $n_D$ is a feature of all the 6 effective excluded-volume coefficients $\widetilde{E}_{1i}$, their rate of change is not the same. Let us focus on the coefficients corresponding to the two-particle configurations (1,4) and (1,5). In absence of depletant ($n_D=0$), $\widetilde{E}_{15}=E_{15}$ is slightly bigger than $\widetilde{E}_{14}=E_{14}$, but its first derivative at $n_D>0$ is smaller. As a consequence, there exists a value of the depletant density $n_D^*$, such that $\widetilde{E}_{14}<\widetilde{E}_{15}$ for $n_D<n_D^*$, and $\widetilde{E}_{14}>\widetilde{E}_{15}$ for $n_D>n_D^*$ (see inset of Fig. \ref{fig2}). This fact, which we will show to have deep consequences for the phase behavior of the system, is more clearly represented by the red solid curve of Fig. \ref{fig3}, representing the ratio $\widetilde{E}_{14}/\widetilde{E}_{15}$ as a function of $n_D$. In the same plot, we report the ratio $\widetilde{E}_{14}/\widetilde{E}_{15}$ relative to boards-cubes mixtures with fixed ratio $w/t=3.0$ and $d/t=1.0$, but different values of $l/t$ ($l/t=8.7$, $9.0$ and $10.0$). In the four cases we observe a monotonically increasing dependence of $\widetilde{E}_{14}/\widetilde{E}_{15}$ on $n_D$, which implies that the existence of $n_D^*$, defined by the condition $\widetilde{E}_{14}=\widetilde{E}_{15}$, is determined by the value of $E_{14}/E_{15}$, which in turn depends only on $l$, $w$ and $t$. In other words, $n_D^*$ exists only if $E_{14} \leq E_{15}$, with $n_D^*=0$ if $E_{14} = E_{15}$. On the contrary, if $E_{14}>E_{15}$ one has $\widetilde{E}_{14}>\widetilde{E}_{15}$ independently of the depletant density $n_D$.

Before addressing the physical consequences of the existence of the density $n_D^*$, it is worth seeing how the relative value of the excluded volume coefficients $E_{14}$ and $E_{15}$ determines the phase behavior of boardlike particles in absence of depletant (i.e. $n_D=0$). It is well-known that monodisperse hard boardlike particles are expected to undergo an $I N$ transition, where the particles spontaneously break the orientational symmetry by aligning along common directions in space \cite{straley}. The nematic phase emerging from the $I$ can be i) {\em uniaxial prolate} $N_+$ with alignment of the long axis $l$; ii) {\em uniaxial oblate} $N_-$ with alignment of the short axis $t$; iii) {\em biaxial} $N_B$ with alignment of the three axes of the particle. Following Onsager \cite{onsager}, the origin of this phase transition can be understood by considering that orientational ordering determines an increase in excluded-volume entropy, which compensates the decrease in orientational entropy. Therefore at fixed orientational entropy, when $E_{14}<E_{15}$ the $N_+$ phase will be thermodynamically favored over the $N_-$, the opposite being the case when $E_{14}>E_{15}$ (cf. Fig. \ref{fig1}). In the intermediate situation, when $E_{14}=E_{15}$, the system undergoes instead a direct second-order $I N_B$ transition. By explicitly calculating $E_{14}$ and $E_{15}$ in terms of $l$, $w$ and $t$, and defining a shape parameter $\nu = l/w-w/t$, one can show that 

\begin{equation}
\frac{E_{14}}{E_{15}}
\begin{cases}
<1 \hspace{0.4cm} \Leftrightarrow \nu>0, \\
=1 \hspace{0.4cm} \Leftrightarrow \nu=0,\\
>1 \hspace{0.4cm} \Leftrightarrow \nu<0.
\end{cases} \label{eq16}
\end{equation}
This is consistent with Straley's result that a system of boardlike particles undergoes i) a first-order $I N_+$ transition if $\nu>0$; ii) a first-order $I N_-$ transition if $\nu<0$; iii) a second-order $I N_B$ transition if $\nu=0$ \cite{mulder}.

\begin{figure*}
\includegraphics[scale=0.7]{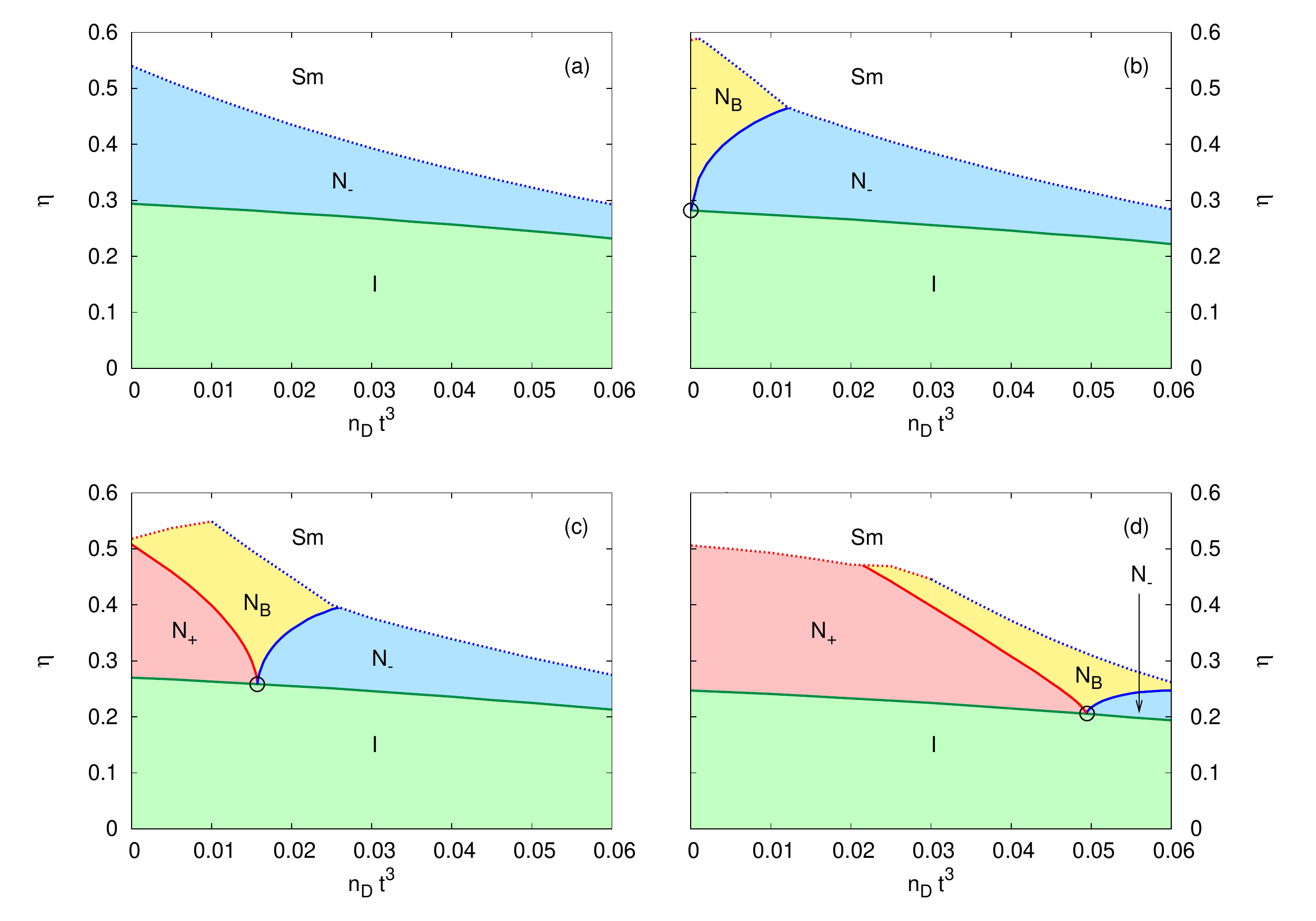}
\caption{\label{fig4} Phase diagrams of boardlike particles with aspect ratios $w/t=3.0$ and (a) $l/t=8.7$ ($\nu=-0.1$), (b) $l/t=9.0$ ($\nu=0.0$), (c) $l/t=9.3$ ($\nu=0.1$), (d) $l/t=10.0$ ($\nu=0.33$) in contact with a reservoir of cubic depletant with side length $d/t=1.0$ at number density $n_D$. The diagrams feature isotropic ($I$, green regions), prolate ($N_+$, red regions) and oblate ($N_-$, oblate regions) uniaxial and biaxial ($N_B$, yellow regions) nematic phases. The black circles highlight the Landau critical points, whereas the dotted lines indicate the limit of stability of nematic phases with respect to smectic ($Sm$) fluctuations along the long (red dotted line) and short (blue dotted line) particle axis respectively.} 
\end{figure*}

The relation between the excluded-volume coefficients and the character of the $I N$ transition can be generalized to the case of boardlike particles immersed in a depletant, provided that the coefficients $E_{i i'}$ are substituted with $\widetilde{E}_{i i'}$. According to this interpretation, one is lead to identify the density $n_D^*$, defined by the condition $\widetilde{E}_{14}=\widetilde{E}_{15}$, with a Landau critical point at which the system undergoes a direct second-order $I N_B$ transition. Since the $N_B$ stability is generally due to a balanced competition between rodlike and platelike behavior, the {\em critical depletant density} $n_D^*$ is expected to divide the phase diagram into two distinct regions, one where the stable uniaxial nematic phase is prolate ($N_+$, corresponding to rodlike behavior), the others where the stable uniaxial nematic phase is oblate ($N_-$, corresponding to platelike behavior).

This picture is confirmed by the $(\eta,n_D)$ phase diagrams of Fig. \ref{fig4}, describing the phase behavior of boardlike particles with dimensions $w/t=3.0$ and (a) $l/t=8.7$ ($\nu=-0.1$), (b) $l/t=9.0$ ($\nu=0.0$), (c) $l/t=9.3$ ($\nu=0.1$), and (d) $l/t=10.0$ ($\nu=0.33$) immersed in a cubic depletant with sides $d/t=1.0$ and number density $n_D$. As a general feature, at packing $\eta \approx 0.2-0.3$ the system undergoes a phase transition from an  $I$ phase (green region) to $N_+$ (red regions), $N_-$ (blue regions) or $N_B$ (yellow regions) states. The first-order character of the $I N_+$ and $I N_-$ transitions is not visible on the scale of Fig. \ref{fig4}. The dotted lines indicate the limit of stability of the homogeneous phases with respect to one-dimensional (smectic, $Sm$) fluctuations along the long axis $l$ (red dotted lines) or along the short axis $t$ (blue dotted lines). Therefore, the smectic bifurcation analysis confirms that inhomogeneous phases (white regions) preempt nematic states at sufficiently high packing fractions. In particular, the higher the depletant density $n_D$, the lower the stability of homogeneous phases with respect to inhomogeneous one. This result is in agreement with previous studies on the phase behavior of hard rods interacting via an attractive depletion potential \cite{bolhuis, savenko}. In the latter case, in fact, the coexistence regions between different phases increase with the depletant density leading, at sufficiently high $n_D$, to a wide isotropic-crystal coexistence and a consequent disappearance of the liquid crystal phases. We expect similar phenomena at depletant concentrations higher than those considered here, but a description beyond the second  virial order would be needed in this case.

As deduced by the analysis of Fig. \ref{fig3}, in the case of ``platelike'' boards with $\nu<0$ (Fig. \ref{fig4}(a)) the absence of a Landau critical point implies that the $I$ phase undergoes a transition towards a $N_-$ phase for every value of $n_D$. If we instead consider a system of boardlike particles with $\nu=0$ (Fig. \ref{fig4}(b)), it is well known that in absence of depletant a direct second-order $I N_B$ transition is expected. In our picture, this corresponds to the presence of a critical depletant density at $n_D^*=0$, beyond which $N_-$ states appear at intermediate packing between the $I$ and the $N_B$ phase. More interesting conclusions can be drawn when considering ``rodlike'' boards with $\nu>0$, in which case the critical depletant density $n_D^*$ assumes non-zero values (Figs. \ref{fig4}(c) and (d)). If the pure system at $n_D=0$ is expected to develop an $I N_+$ transition, the attraction induced by the depletant reduces the $N_+$ stability until determining a direct $I N_B$ transition at $n_D=n_D^*$. Surprisingly, at even higher depletant densities ($n_D>n_D^*$), the stable uniaxial nematic phase in between $I$ and $N_B$ has oblate ordering $N_-$, in sharp contrast with the behavior of the pure rodlike boards system. Moreover, the phase diagrams of Figs. \ref{fig4}(c) and (d) suggest that, when dealing with boardlike particles with $\nu>0$, setting the depletant density at values close to $n_D^*$ allows to select regions of the phase diagram with relatively high $N_B$ stability. This is possible also when the regime of $N_B$ stability of the pure boardlike particle system is small (Fig. \ref{fig4}(c)) or even absent (Fig. \ref{fig4}(d)).

A relevant feature of the present analysis is that a critical depletant density $n_D^*$ exists only for slightly elongated, or rodlike, boards ($\nu > 0$). On the contrary, no Landau critical point is predicted when $\nu<0$, in which case for every value of $n_D$ the system develops a first-order $I N_-$ transition, typical of platelike particles. This fact can be interpreted in the following terms. At low enough depletant density ($n_D<n_D^*$) the role of the depletant is weak and the isotropic-nematic phase transition is driven by the gain in boardlike particles' excluded-volume entropy, leading to $N_+$ ($N_-$) ordering if $\nu>0$ ($\nu<0$). On the other hand, at high enough depletant density ($n_D>n_D^*$) the thermodynamically more favored states are those maximizing the depletant entropy, i.e. states where the overall depletion volume is minimized. It appears clear that, at fixed boardlike particles' orientational entropy and independently of the sign of $\nu$, $N_-$ rather than $N_+$ ordering tends to maximize the overlap between the depletion regions of single boards. Therefore, when $\nu>0$ the Landau critical point at $n_D^*$ appears as a result of a competition between the excluded-volume entropy of boardlike particles and depletant. Instead, when $\nu<0$ this competition does not happen since both entropies are maximized by $N_-$ states, and thus no critical depletant density exists.

\begin{figure}
\includegraphics[scale=0.65]{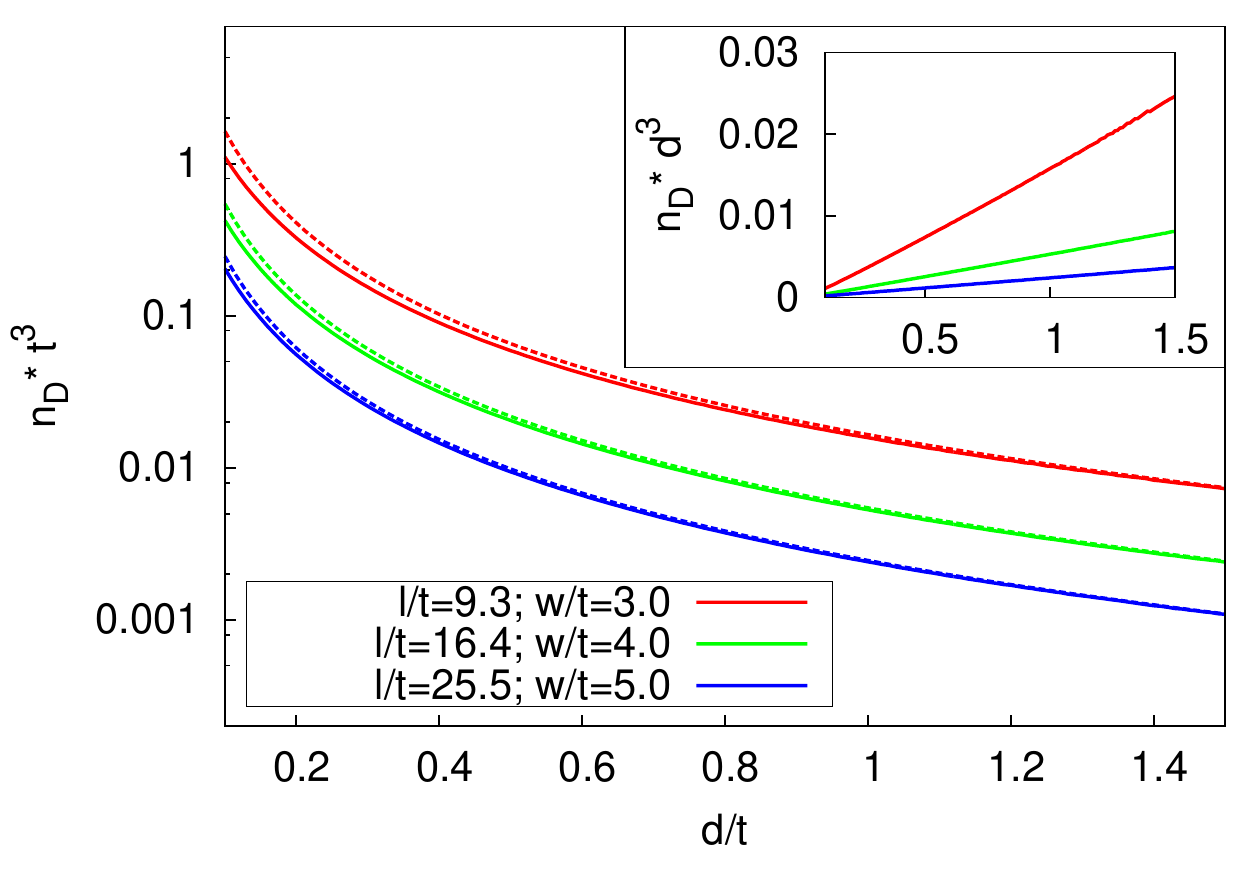}
\caption{\label{fig5} Critical depletant density $n_D^*$ as a function of the side $d$ of the cubic depletant for different boardlike particles with same shape parameter $\nu=0.1$: $l/t=9.3$ and $w/t=3.0$ (solid red line); $l/t=16.4$ and $w/t=4.0$ (solid green line); $l/t=25.5$ and $w/t=5.0$ (solid blue line). The dashed lines represent the approximate analytical dependence given by Eq. (\ref{eq17}). The inset illustrates the same data in terms of the critical depletant packing fraction $n_D^* d^3$.} 
\end{figure}

One could wonder how the phase behavior of our system changes by varying its relevant parameters, that is, the dimensions of the boardlike particles $l$, $w$, $t$ and the size of the depletant $d$. The phase diagrams of Fig. \ref{fig4} highlight the monotonic increasing dependence of $n_D^*$ on $\nu$, when the dimensions of the depletant $d/t$ and the aspect ratios of one of the two boards (here $w/t$) are fixed. By numerically solving the equation $\widetilde{E}_{14}=\widetilde{E}_{15}$, we investigate also the role of the depletant dimension on the critical depletant density. In Fig. \ref{fig5} we report the critical depletant density $n_D^*$ as a function of the depletant side $d/t$ for three boardlike particle dimensions with different aspect ratios $l/t$ and $w/t$, but same shape parameter $\nu=l/w-w/t=0.1$. As a general trend, by increasing the size of the depletant, the critical depletant number density $n_D^*$ decreases, in accordance with the intuitive notion that the smaller the depletant, the more one needs to establish enough depletion attraction. Moreover, at fixed $\nu$ the critical depletion density decreases most for the more extreme aspect ratios of the particles. In other words, the bigger the aspect ratios $l/t$ and $w/t$ at fixed $\nu$, the smaller the amount of depletant needed to reach $n_D^*$. If instead of the number density one considers the critical depletant packing fraction $n_D^* d^3$, one sees that this quantity is an increasing function of $d$ (inset of Fig. \ref{fig5}). Therefore, one expects the ideal depletant approximation (cf. Sec. \ref{sec:model}) to be increasingly reliable in the limit of small depletant.

In practical situations, one could be interested in estimating the critical depletant density $n_D^*$, which is defined as the solution of the non-linear equation $\widetilde{E}_{14}=\widetilde{E}_{15}$ and which has then to be calculated numerically. If $n_D^*$ is sufficiently small, one can obtain an approximate expression for this quantity by linearizing both sides of equation $\widetilde{E}_{14}=\widetilde{E}_{15}$ in $n_D$. The approximate critical depletant density is given by the following expression

\begin{equation}
n_D^* = \frac{2 \bigl [t(l+w)^2 - l(w+t)^2 \bigr](l-t)^{-1}}{2(lt-w^2) d^3 + w (lw+tw-2lt) d^2}, 
\label{eq17}
\end{equation}
and it can be compared (dotted lines of Fig. \ref{fig5}) with the numerical calculation (solid lines), showing good overall agreement, which improves the larger the depletant side $d/t$.

\section{Conclusions}
\label{sec:conclusions}

In the present paper we investigate for the first time the effect of a short-range depletion-induced attraction on the liquid-crystal phase behavior of boardlike particles. To this aim, we make use of classical density functional theory truncated at second virial order, and adopt the Zwanzig model for the description of the orientational degrees of freedom. In close analogy with the Asakura-Oosawa-Vrij model for mixtures of spheres, by neglecting interactions between the cubic depletant particles we can explicitely calculate the effective two-body attractive depletion potential between boardlike particles.

We predict that in systems of slightly elongated boardlike particles ($\nu>0$), there exists a critical depletant density at which the uniaxial nematic phase is substituted by a direct second-order transition from the isotropic to a biaxial nematic phase. At higher depletant concentrations, a large region of oblate uniaxial nematic ordering develops, rendering the system of attractive rodlike boards behaving like a system of hardly repulsive platelike boards. The origin of this phenomenon is due to two competing mechanisms: the maximization of the boardlike particle entropy, favoring $N_+$ ordering, and the maximization of the depletant entropy, favoring $N_-$ ordering. 

The phase behavior described in this work shares many similarities with our findings in Ref. \cite{belli}, where we showed that size-polydispersity in a system of hard boardlike particles with the same shape and different volume induces the appearance of a Landau tetracritical point at a specific system composition. This fact is related to a competition between prolate and oblate ordering, which in turn is realized only when the boardlike particles are slightly elongated. In the light of our present findings, we suggest that this $N_+$-$N_-$ competition and the corresponding emergence of a Landau tetracritical point can be understood in terms of a depletion effect. More specifically, when size-polydispersity becomes relevant, $N_-$ rather than $N_+$ ordering determines the higher total entropy due to the minimization of the overall depletion regions of the big particles with respect to the smaller ones. In further analogy with the boards-depletant mixture, no such competition is predicted for platelike boards and consequently no tetracritical point appears in this case.

Besides furnishing an explanation for the results of Ref. \cite{belli}, we suggest that manipulating the attraction induced by a depletant, e.g. a non-adsorbing polymer, furnishes an original and effective way to control the phase behavior of boardlike particles, allowing to stabilize prolate and oblate uniaxial and biaxial nematic states. Moreover, the depletant density is expected to be an easy experimental control parameter.

\section{Acknowledgements}

It is our pleasure to thank R. Kamien for suggesting, during the 8th Liquid Matter Conference in Vienna, to interpret the results of Ref. \cite{belli} as a depletion effect. We would also like to thank R. Ni for stimulating discussions.

This work is financed by an NWO-VICI grant and is part of the research program of the ``Stichting door Fundamenteel Onderzoek der Materie (FOM)'', which is financially supported by the ``Nederlandse Organisatie voor Wetenschappelijk Onderzoek (NWO)''.

\end{document}